\documentclass[11pt,a4paper]{article}

\usepackage{amsmath,amssymb,graphicx,cite}

\def\lp{\left(}
\def\rp{\right)}
\def\lb{\left[}
\def\rb{\right]}
\def\be{\begin{equation}}
\def\ee{\end{equation}}

\begin{document}

\title{Thin shells in Einstein--Born--Infeld theory} 
\author{Ernesto F. Eiroa$^{1,2,}$\thanks{e-mail: eiroa@iafe.uba.ar}, 
Claudio Simeone$^{2,3,}$\thanks{e-mail: csimeone@df.uba.ar}\\
{\small $^1$ Instituto de Astronom\'{\i}a y F\'{\i}sica del Espacio, C.C. 67, Suc. 28, 1428, Buenos Aires, Argentina}\\
{\small $^2$ Departamento de F\'{\i}sica, Facultad de Ciencias Exactas y 
Naturales,} \\ 
{\small Universidad de Buenos Aires, Ciudad Universitaria Pab. I, 1428, 
Buenos Aires, Argentina} \\
{\small $^3$ IFIBA, CONICET, Ciudad Universitaria Pab. I, 1428, 
Buenos Aires, Argentina}} 
\maketitle

\begin{abstract}

The characterization and mechanical stability of charged thin shells with spherical symmetry are analyzed in the context of Einstein-Born-Infeld theory. The study of stability is performed by considering linearized perturbations preserving the symmetry of the static configurations. It is found that as the charge increases, the shells can be stable for a wider range of the parameters. 




\end{abstract}

\section{Introduction}

Born and Infeld introduced a nonlinear theory of the electromagnetic field \cite{borninf} with the aim to solve the problem of the infinite self energy  of a charged point particle in Maxwell electrodynamics. The  spherically symmetric solution for Einstein gravity coupled to Born--Infeld electrodynamics was first obtained by Hoffmann \cite{hoffmann}. This solution does not describe the electron, but it corresponds to a black hole. The field equations of Einstein gravity coupled to Born--Infeld electrodynamics have the vacuum spherically symmetric solution \cite{gibbons,breton} 
\be
ds^{2}=-f(r)dt^{2}+f^{-1}(r)dr^{2}+r^{2}(d\theta^2+\sin^2\theta d\varphi^2),
\label{bi1}
\ee
with
\be
f(r)=1-\frac{2M}{r}+\frac{2}{3b^{2}}\left\{ r^{2}-\sqrt{r^{4}+b^{2}Q^{2}}+
\frac{\sqrt{|bQ|^{3}}}{r}F\left[ \arccos\left( \frac{r^{2}-|bQ|}{r^{2}+|bQ|}
\right) ,\frac{\sqrt{2}}{2}\right]\right\} ,
\label{bi2}
\ee 
where $F(\gamma ,k)$ is the elliptic integral of the first kind. As usual, $M>0$ is the ADM mass and $Q^{2}=Q_{E}^{2}+Q_{M}^{2}$ is the square of the charge. The corresponding electric and magnetic inductions are $D(r)=Q_{E}/r^{2}$ and $B(r)=Q_{M}\sin \theta$. With the units adopted ($c=G=1$), $M$, $Q$ and $b$ have dimensions of length. The parameter $b$ measures the departure from Maxwell theory. In the limit $b\rightarrow 0$, the Reissner--Nordstr\"{o}m metric $f(r)=1-2M/r+Q^2/r^2$ is obtained. The geometry given by Eqs. (\ref{bi1}) and (\ref{bi2}) is also asymptotically Reissner--Nordstr\"{o}m for large values of $r$ and it is singular at the origin \cite{breton}. The zeros of $f(r)$ correspond to the horizons. For a given value of $b$, when $0\le |Q|\le Q_d$, the function $f(r)$ has only one zero corresponding to a regular event horizon. For $Q_d<|Q|<Q_c$, $f(r)$ has two zeros; then, an inner horizon and an outer regular event horizon exist. When $|Q|=Q_c$, there is one degenerate horizon. Finally, if $|Q|>Q_c$, the function $f(r)$ has no zeros and there is a  naked singularity. The values of $|Q|$ where the number of horizons change, which result from the condition $f(r_{h})=f'(r_{h})=0$, are increasing functions of $|b|/M$. 

The Darmois--Israel formalism \cite{daris}, in which the Lanczos equations relate the geometry at both sides of a surface with the induced energy-momentum tensor on it, has become the main tool for the study of the dynamics of highly symmetric layers. The stability of spherical shells  was studied by several authors (see  \cite{poi,sh1,sh2,sh3} and the references included there). The formalism was applied to bubbles, shells around stars and black holes, and to thin-shell wormholes (see for example \cite{wh1,wh2,wh4} and references therein). Here we present a recent study \cite{eisi11} of the mechanical stability of spherical shells under perturbations preserving the symmetry within the framework of Einstein--Born--Infeld theory. We first introduce the formalism and perform the mathematical construction of shells. Then, we analyze charged bubbles and charged layers around non charged black holes in the framework of Einstein gravity coupled to Born--Infeld electrodynamics.

\section{Charged shells: construction and stability}

We start from two manifolds $\mathcal{M}_1$ and $\mathcal{M}_2$ with metrics of the form (\ref{bi1}) in  coordinates $X_{1,2}^\mu=(t_{1,2},r_{1,2}, \theta, \varphi)$, with in general different arbitrary functions  $f_{1,2}$, respectively. We cut and paste them at the spherical surface $\Sigma$ defined by $r_{1,2}=a$. The resulting manifold  ${\cal M}$ is given by the union of the inner part ($r_1\leq a$) of $\mathcal{M}_1$ and by the outer part ($r_2\geq a$) of $\mathcal{M}_2$ . The line element is continuous across $\Sigma$ if the coordinates at each side are set to satisfy $f_1(a)dt_1^2=f_2(a)dt_2^2$, as required by the thin-shell formalism \cite{daris}. We let the radius $a$ to be a function of the proper time $\tau$ measured on the surface. The coordinates at the surface $\Sigma$ are $\xi ^i=(\tau , \theta,\varphi )$. The relation between the geometry and the shell matter is given \cite{daris} by the Lanczos equations  
\be
-[K_{ij}]+g_{ij}[K]=8\pi S_{ij},
\label{leq}
\ee
where $g_{ij}$ is the induced metric on $\Sigma$, $K_{ij}$ is the extrinsic curvature, $K$ is its trace, and $S_{ij}$ is the surface energy-momentum tensor; the brackets denote the jump of a given quantity $q$ across the surface: $[q]=q^2\vert_\Sigma -q^1\vert_\Sigma$. If  $[K_{ij}]\neq 0$ we have a thin shell at $\Sigma$.  The Lanczos equations  give the energy density and the pressure at the shell
\be
\sigma=-S_\tau^\tau= -\frac{1}{4\pi a}\lp\sqrt{f_2(a)+{\dot a}^2}-\sqrt{f_1(a)+{\dot a}^2}\rp,\label{00}
\ee
\be
p=S_\theta^\theta=S_\varphi^\varphi=-\frac{\sigma}{2}+\frac{1}{16\pi}\lp\frac{2\ddot a+f'_2(a)}{\sqrt{f_2(a)+{\dot a}^2}}-\frac{2\ddot a+f'_1(a)}{\sqrt{f_1(a)+{\dot a}^2}}\rp,
\label{pres}
\ee
where a prime represents  $d/dr$, and a dot stands for $d/d\tau $. The equations above, or any of them plus the conservation equation
$d(a^2\sigma)/d\tau+p da^2/d\tau=0$, determine the evolution of the shell.  We consider small perturbations preserving the symmetry around a static solution. Our procedure is similar to the treatment in Refs. \cite{poi,sh1,sh2,sh3,wh1}. Provided the equation of state $p=p(\sigma)$, the conservation equation can be formally integrated \cite{wh1} to give $\sigma=\sigma(a)$.  From Eq. (\ref{00}) we obtain
\be
{\dot a}^2+V(a)=0,
\label{energy}
\ee 
where 
\be
V(a)=S-\frac{1}{4}\lp \frac {m}{a}\rp^2-\lp \frac{a}{m}\rp^2 R^2 ,\label{potential}
 \ee 
with $S(a)=\lp f_1(a)+f_2(a)\rp /2$, $R(a)=\lp f_1(a)-f_2(a)\rp /2$ and $m(a)=4\pi a^2\sigma$. An equilibrium radius $a_0$ satisfies $V(a_0)=0$ and $V'(a_0)=0$, and stability requires $V''(a_0)>0$. After evaluating the derivatives, the condition for stability gives 
\be
\frac{m}{4a_0}\lp\frac{m}{a_0}\rp''+ \frac{a_0}{m}\lp\frac{a_0}{m}\rp''R^2 < \Omega(a_0)-\Gamma^2(a_0),
 \ee 
where
\be
\Gamma(a_0) = \frac{a_0}{m}\lb S'-2\frac{a_0}{m}\lp\frac{a_0}{m}\rp'R^2-2\lp\frac{a_0}{m}\rp^2 RR'\rb
\ee
and
\be
\Omega(a_0)  =  \frac{S''}{2}-\lb\lp\frac{a_0}{m}\rp'\rb^2R^2 -4\frac{a_0}{m}\lp\frac{a_0}{m}\rp'RR' - \lp\frac{a_0}{m}\rp^2\lb R'^2+RR''\rb.
\ee
Here $m$, $S$ and $R$ are given as functions of $a_0$. Using the conservation equation and defining $\eta = p'(a_0)/\sigma'(a_0)$ the condition for stable equilibrium can be put in the form
\be
2\lp \frac{a_0}{m} \rp ^4 \lb \lp\frac{m}{a_0}\rp' \rb ^2 R^2 + \frac{1}{a_0}\lb \frac{m}{4a_0} - \lp \frac{a_0}{m}\rp ^3 R^2 \rb \lb \frac{m}{a_0^2}-\lp\frac{m}{a_0}\rp'\rb \lp 1+2\eta \rp<\Omega(a_0)-\Gamma^2(a_0).
\label{sta}
\ee
The subsequent study is carried out in terms of the parameter $\eta $, which for $0<\eta\leq 1$ can be understood as the square of the velocity of sound on the shell. The results above are also valid for wormholes if the outer parts of $\mathcal{M}_1$ and $\mathcal{M}_2$ are taken, and the $(-)$ sign inside the parenthesis in Eqs. (\ref{00}) and (\ref{pres}) is changed by a $(+)$ sign. In particular, if $\mathcal{M}_{1,2}$ are equal copies of the geometry (\ref{bi1}) with $r\geq a$ the results of Refs. \cite{wh1,wh2,wh4,risi09} can be recovered.

\begin{figure}[t!]
\centering
\begin{minipage}{.47\textwidth}
\centering
\includegraphics[width=\textwidth]{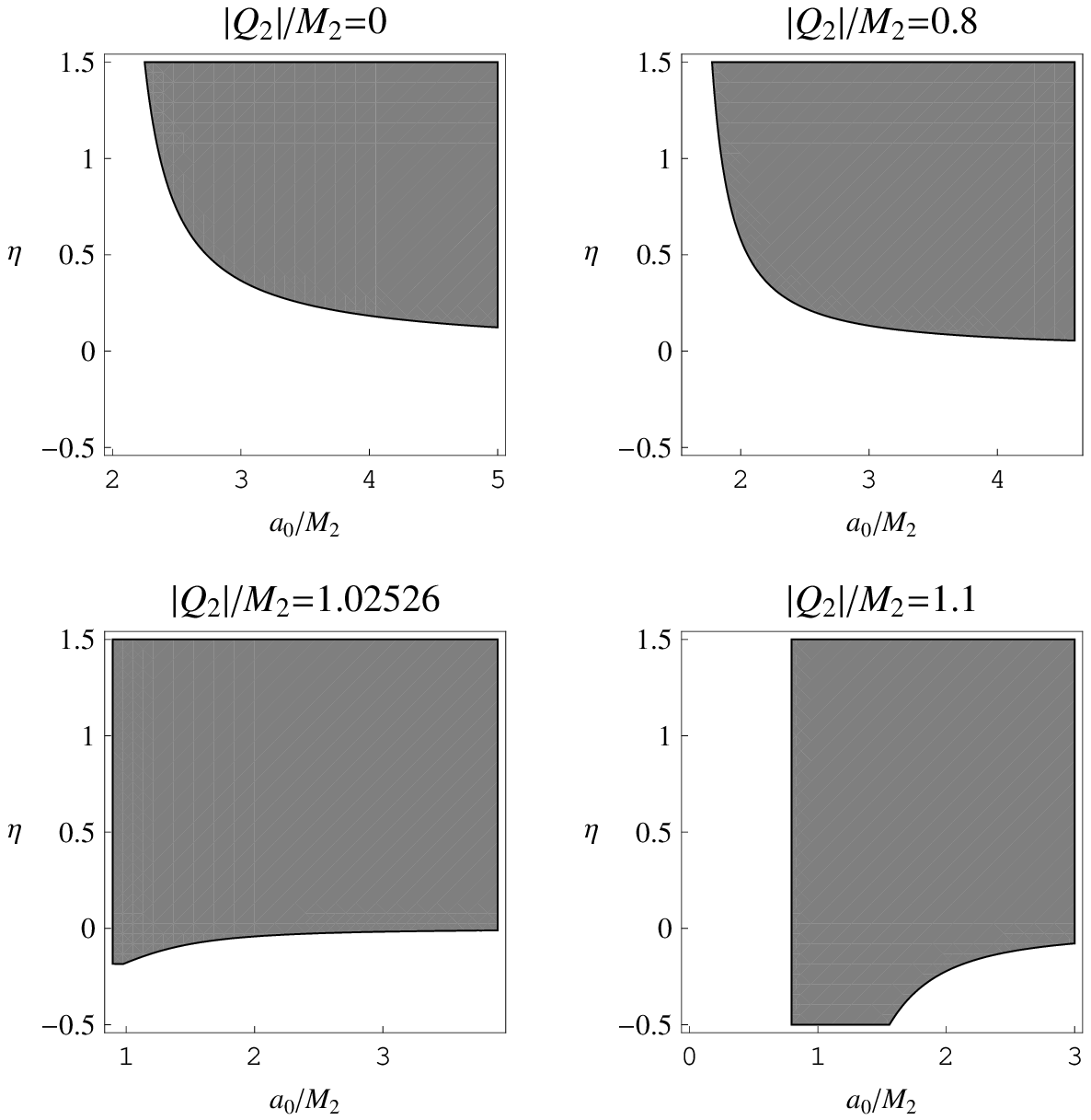}
\caption{Stability regions (grey) for charged shells of normal matter around vacuum ($M_1=0$, $Q_1=0$).}
\label{fig1}
\end{minipage}
\hfill
\begin{minipage}{.47\textwidth}
\centering
\includegraphics[width=\textwidth]{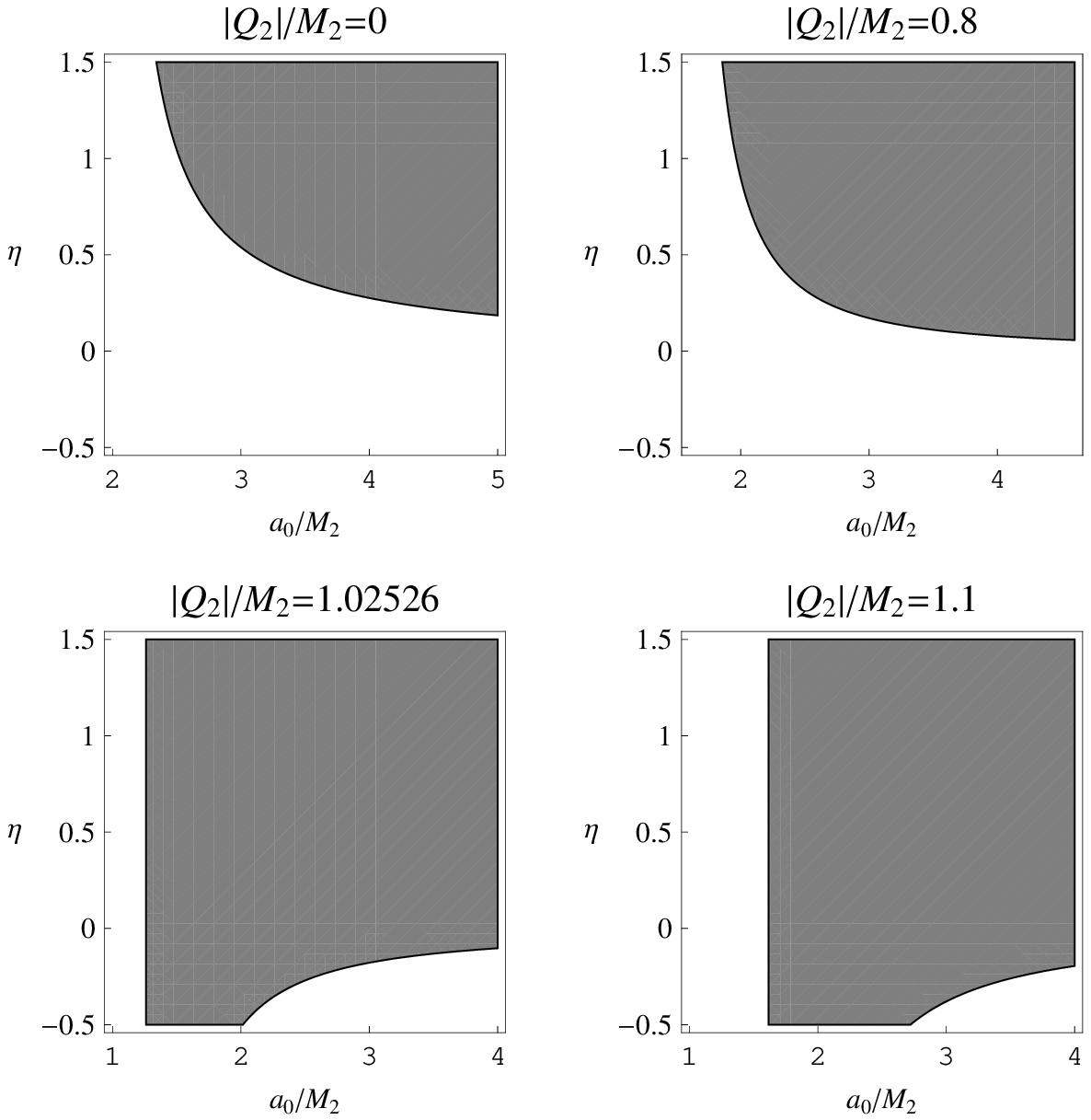}
\caption{Stability regions (grey) for charged shells of normal matter around a black hole ($M_1/M_2=0.5$, $Q_1=0$).}
\label{fig2}
\end{minipage}
\end{figure}

The geometry consists in the inner part of manifold $\mathcal{M}_1$ and the outer part of manifold $\mathcal{M}_2$ joined by the shell $\Sigma $. Our analysis is restricted to normal matter, so that the weak energy  condition $\sigma\geq 0$ and $\sigma+p\geq 0$ must be fulfilled. We first consider bubbles, so  we take the outer manifold with mass $M_2$ and charge $Q_2$, and the inner one with both vanishing mass and charge. The radius $a_0$ is chosen larger than the horizon radius of the outer manifold (so that the singularity and the event horizon of the original manifold are both removed). In second place, we analyze charged shells around non charged black holes, so $M_1\neq 0$, $Q_1=0$, $M_2\neq 0$ and $Q_2\neq 0$. Besides demanding that $a_0$ is greater than the horizon radius of the outer manifold, we also have to take $a_0>2M_1$. A necessary condition (but not sufficient for charged shells) for fulfilling the weak energy condition is that $0\leq M_1<M_2$. We present the results graphically in Figs. \ref{fig1} and \ref{fig2}; we take $b=1$, which means a large deviation from Maxwell electrodynamics. We do not restrict our analysis to the range $0<\eta \le 1$, in which $\eta^{1/2}$ can be interpreted  as the velocity of sound on the shell, though the results within this range are of more physical interest. The figures illustrate the dependence of the stability regions in terms of the parameters and the constant $b$; in particular, for shells surrounding black holes we show the results corresponding to $M_1/M_2=0.5$. For comparison, the non charged shells results of Ref. \cite{poi} are also displayed. The qualitative behaviour is different if the charge is under or beyond the critical value $Q_c$ (from which the horizon of the outer original manifold vanishes). If $Q<Q_c$, stable configurations are possible only for positive $\eta $, while if $Q\geq Q_c$, negative values of $\eta $ are compatible with stability. In all cases, there are values of $a_0/M_2$ for which stable configurations are possible with $0<\eta \leq 1$.  Within this range of $\eta $, if the charge is below $Q_c$ the largest interval of $a_0/M_2$ for which the shell is stable corresponds to $\eta =1$, as it was obtained for non charged shells in Ref. \cite{poi}. When the charge is under $Q_c$, shells around black holes present slightly smaller regions of stability than bubbles; and the stability regions become larger as the charge increases (the same happens for spherical thin-shell wormholes with charge \cite{wh2}). If the charge is equal or beyond $Q_c$, for a given $\eta$ bubbles can be stable for smaller radii $a_0/M_2$ than shells around black holes, and for fixed $a_0/M_2$ bubbles are stable for a smaller range of the parameter $\eta $ than shells around black holes.

\section{Summary} 

We have studied the stability under perturbations preserving the  symmetry for bubbles and shells around non charged black holes within the framework of Einstein--Born--Infeld theory. The presence of the charge seems to enlarge the stability regions  for both bubbles and shells around black holes. The stability regions for bubbles appear to be larger than those of shells around black holes if the charge is under the critical value, while the reverse is true for charges  above the critical value.  Charged layers with $0<\eta\leq 1$ can be stable for suitable values of the parameters. From a different point of view, we note that with small changes in the formalism the stability of thin-shell wormholes in the same theoretical framework has also been studied (see Ref. \cite{risi09}).

\section*{Acknowledgments}

This work has been supported by Universidad de Buenos Aires and CONICET.

\end{document}